# New framework model to secure Cloud data storage

Leila Beldjezzar
LIRE Labs
University of Constantine 2
25000 Constantine, Algeria
Megouache_leila @yahoo.fr

Abdelhafid Zitouni
LIRE Labs
University of Constantine 2
25000 Constantine, Algeria
azitouni@yahoo.fr

Mahieddine Djoudi
TechNE Labs
University of Poitiers
86073 Poitiers Cedex, France
mdjoudi@univ-poitiers.fr

*Abstract*—Nowadays companies are increasingly adopting the technology of cloud computing. This technology is subject to a lot of research and continuous advances are made. The use of cloud computing in the companies advantages such as: reducing costs, sharing and exchange of information between institutions, but the data in the Cloud computing are susceptible to be compromised and the companies are exposing to see their data loss. In this study, we address the subject of security in cloud computing; we expose and discuss some researches that had been proposed to secure the data stored in the cloud. And then we will present our new frameworks that ensure confidentiality of data storage in the cloud environment.

*Keywords*— *Cloud computing; performance; availability; security; authentication.*

## I. INTRODUCTION

Cloud computing is one of the new technologies that appeared in these last years. Its main objectives are to deliver different services for users, such as infrastructure, platform or software with a reasonable and more and more decreasing cost for the users. However, cloud computing is still in its initial stage [1]. The lack of standards, the security and the interoperability issues hamper the growth of cloud computing [2, 3]. Thus, the choice of clouds made by companies is usually based on the quality of services, but measuring the quality of cloud providers' approach to security is difficult because many cloud providers will not expose their infrastructure to customers.

Therefore, the security is an important factor that should be taken into account by cloud service providers.

Before making the transfer of the data towards Cloud, the company owes classify them and choose Cloud adapted according to the following categories.

The deployment models in cloud computing are:

a) Public cloud: the service is provided by a third party via the internet. The physical infrastructure is owned and managed by the service provider. This cloud is less secure compared to other models [4], since all applications and data are available to the public and accessible via the Internet.

b) Private cloud: it is a dedicated cloud that is managed internally or by a third-party and can be hosted generally on premises or even externally [5]. The physical infrastructure is exclusively used by one organization. This Cloud offer a higher degree of security since only users in the organization have access to the private cloud.

c) Community cloud: the physical infrastructure is controlled and shared by several organizations and is based on a community of interest [6].

d) Hybrid cloud: combine two or more distinct cloud infrastructures. This cloud must be linked by a standard technology for data and applications portability.

The main characteristics of Cloud computing are [6]:

a) On-demand self-service: users can access and control their services automatically without the intervention of the service provider.

b) Broad network access: Services are available over the internet, they accessible from anywhere regardless of the device used.

c) Resource pooling: the computing resources are pooled to serve multiple consumers, with different physical and virtual resources [6].

d) Rapid elasticity: The computing resources are rapidly and dynamically provisioned. Users can increase and decrease the functionality available according to their needs. The capacities appear to be unlimited.

e) Measured service: The resource use is controlled by leveraged metering capability.

This paper is organized as follows: Section II introduces cloud computing and cloud security. In section III, the related works is presented. In section IV, we propose our framework. Finally, In Section V, the conclusions of this work are presented.

## II. CLOUD COMPUTING SECURITY

In this section we will present the different security problems in cloud computing such as if the data loss or leakage.

Moving sensitive data to the cloud involves moving the control of the data to the service provider [7]. Therefore, the security and confidentiality of information becomes a major concern data security and authentication in cloud computing is similar to data security and privacy in traditional environments [8].

However, because of the characteristic of opening and multi-location of cloud computing, data security and privacy face to more risks.

The user data need to be protected [6]:

*1) Personally identifiable information:* Includes any information that can be used to identify an individual, such as name, address, Tel...

*2) Sensitive information*: requires additional protection. Such as personal financial information on job performance information and information considered being sensitive personally identifiable information such as biometric information or collections of surveillance camera images in public places

*3) Usage data:* consists of information collected from computer devices such as printers or habits of researchers.

*4) Unique device identities*: information that could be attached to a user device, such as IP address, unique hardware identities, for example the health data is personal and sensitive information.

*5)* **Implementation of a program of data protection** [9].

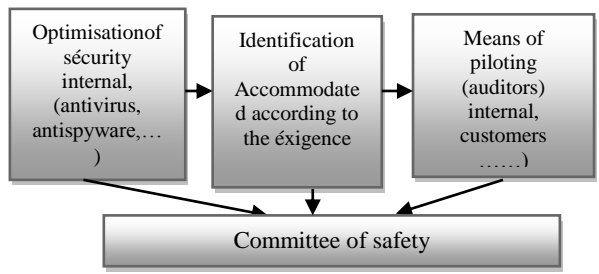

Figure (1): how to secure data

Optimization of security interne by setting up devices internal for protect their systems.

Identification of Accommodated according to the exigency concerns the statutory particular requirements or the jobs which it is necessary to identify [10].

Means of piloting the customer has to have ways of piloting and operational follow-up supplied by the service provider Cloud.

Committee of safety between both parts (customer, supplier of service) where the person receiving benefits creates a " mutualzed and secure environment ".

*6)* **The selection of the offer Cloud:** the most adapted to the need must be made with operational consideration of the precautions and the contract employees.

In terms of protection, Cloud Security Alliance has proposed two action methods [11][12]:

• The first method (and the simplest) is to ensure fine control of data access, fully through identity and access management (e.g. ensuring that the raw data couldn't be accessed by human users, monitoring the access to requesters, authenticating its users).

• If necessary (regulatory perspective), encrypting the most sensitive data. But to be efficient, an encryption solution must appoint means of access control to fine-grained data and encryption key management, while maintaining a high level of performance.

### III. RELATED WORKS

There are a number of work concerning the security, the privacy and the authentication of companies data in the cloud computing.

[8]: Indeed the cloud is a virtual space, which contains data that is fragmented; the data fragments are always duplicated and distributed on physical storage media in addition the cloud contains a restitution function to restore the data. But in this solution granularity of the selected fragmentation is important (the fragments may be too big or too small)

[11]: In this article, the authors present the models to maintain the confidentiality of data handled at the data integration system. The draft PAIRSE addresses the challenge of flexible and preserving privacy in data integration system.
To ensure protection of the data, the authors proposed execution model preserve privacy for data services (data services) that enable service providers to respect their privacy and security policies. The advantage of this model is added only a small increase of the execution time of the service and that model to protect the confidentiality of the data handled in the data integration system.

[12]: Proposes an access control mechanism to ensure confidentiality of data in the cloud. The mechanism is based on two protocols: ABE (Attribute Based Encryption) for data privacy and ABS (Attribute Based Signature) for user authentication. ABE is combined with ABS to ensure anonymity of users that store their data in the cloud. Key attributes and distribution is done in a Decentralized Manner.

[13]: Provides an overview of Common approaches to preserve confidentiality in e-Health Cloud. These approaches are classified into two categories:
Cryptographic approaches (based on encryption techniques) and non-cryptographic approaches (mainly use on access control).They also point out the advantages and disadvantages of each approach.

[14] : In this, the authors present a contribution to protecting the privacy of Web users. The objective of this work is to allow a client to query the search engine in a way to preserve privacy. This means that the search engine, which receives the request, or any opponent who listens to the network, cannot deduce the identity of the applicant (the user). the authors aim to generate false application (Fake query) that cannot be identified by the opponents (or engine research).

A major disadvantage is the introduction of irrelevant answers to protect the applications in this solution because they added noise to the search request to perform obfuscation (interference). This solution decreases the precision of the results and causes overload on the network.

## IV. PROPOSED SECURITY MODEL

Major concerns and issues in security have been discussed in the previous sections. It has been observed that, despite quality research on security data outsourcing and data services for almost a decade, existing approaches on database encryption, certification, digital signatures [12], contractual agreements etc. have not gained much success in operations.

To date, there is minimal work done in the field of security of data as compared to traditional data storage. Different approaches are discussed with assorted categories of confidentiality, privacy, integrity and availability.

Our proposal framework be to create one networks virtual deprived between the customer and the Cloud of such goes out that the customer little to reach his space Cloud in a secure way, to add has it a double cryptography authentication. The figure (2) demonstrates our framework.

A software given to the supplier to the customer called *VPN* (virtual privacy network) *customer,* who allows establishing the connection of the customer his virtual network with a users and a password. This information shall be encrypted automatically.

Our framework consists of five steps, that each has a function explained as follows:

- *Establishment of the connexion*: The customer opens the application which is installed on it computer called "*vpncustomer*". Made a connect, a window opens asking him of entered the user name and the password this password is encrypted.
- *Open the http address*: Where the application will once be connected, the customer goes on the internet page and entered the URL of the cloud; this URL is delivered by the provider.
- *Identification and password:* Once the user name and the word of pass seized for second time, at the same time this information is encrypted by the application according to the algorithm of encryption.
- *Check of the certificate*: The supplier of the cloud Verify has every entered the validity of the right of the customer as well as its certificate which includes everything piece of information of the customer (the last update, the expiry date of its contract…)
- *Secure Access:* The Access of the customer in their space is totally secured.

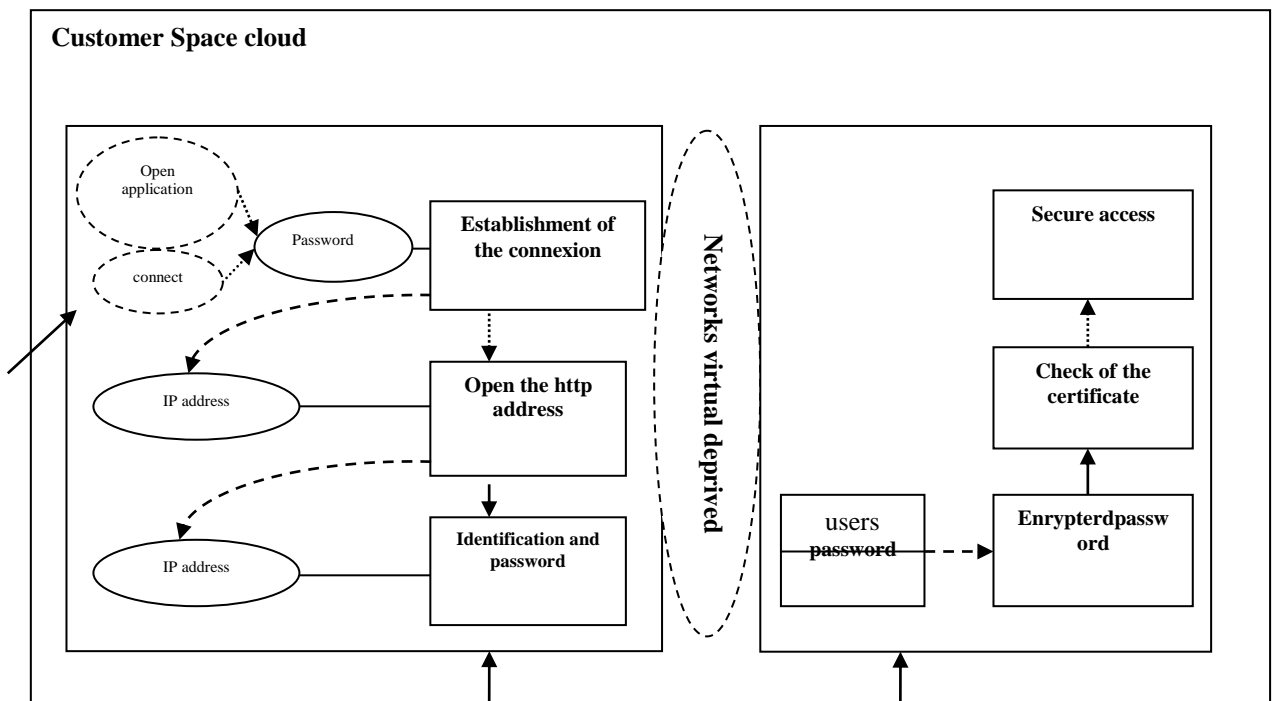

**Figure (2): Prototype of safety**

A software given to the supplier to the customer called **VPN** (virtual privacy network) *customer,* who allows establishing the connection of the customer his virtual network with a users and a password. This information shall be encrypted automatically.

Our framework consists of five steps, that each has a function explained as follows:

Wherever, we will use the public key cryptography just to exchange the symmetric key between the restitution and encryption program. Advanced Encryption Standard or **AES** (**S**ymmetric **E**ncryption **A**lgorithm) is a symmetric encryption algorithm. He won in October 2000 the AES competition, launched in 1997 by the NIST and became the new encryption standard for US Government organizations.

The following procedure describes the overall operation of AES. It takes as input a data table T (clear text) that is modified by the procedure and returned output (ciphertext).

*Input : table **T** and key **K***
*Output : table **T** modified*
*Function AES (**T**, **K**)*
*Begin*
*KeyExpansion (**K**, **TK**);*
*AddRoundKey (T, TK [0];*
*for (i = 1; i<nr; i + +)*
*Round (**T**, **TK** [i]);*
*FInalRound (**T**, **TK** [nr]);*
*End*

**The algorithm used to encrypt data**

*Decryption with AES:*
The encryption routine can be reversed and rearranged to produce a decryption algorithm.
*AES_Decrypt(T, K) {*
*KeyExpansion(K, RoundKeys);*
*/* Initial addition */*
*AddRoundKey(State, RoundKeys[Nr]);*
*for (r=Nr-1; i>0; r--) {*
*InvShiftRows(T);*
*InvSubBytes(T);*
*AddRoundKey(T, RoundKeys[r]);*
*InvMixColumns(T);*
*}*
*/* FinalRound */*
*InvShiftRows(Out);*
*InvSubBytes(Out);*
*AddRoundKey(Out,RoundKeys[0]);*
*}*

**The algorithm used to decrypt data**

When the connection will be establishes, the customer goes on the internet page and opens him on the address URL, Which is delivered by provider, when the page opens another user and password are required **figure (2)** ; the customer little to reach his workspace Cloud, to put this data in the daytime, and to consult them, or used the available applications. The work of the customer will be protected by two authentications in entries and even the little protected customer a copy of its work at his home, and nothing will be lost or destroys even in the case of cut or maintenance.

-Authentication Protocol

This diagram of sequence explains better our architecture which is in the figure 1.Our process follows the following steps:

*Connect;*

**Step 1**: it the phase of registration, the connexion is established between the customer and the private network.
*Customer gOf private netwok*
**Step 2:** the user's name and key is verified (by encrypted and decrypted algorithm).
*Encrypted/decrypted*
https:// adress
**Step 4:** generate Contract**,** in this step the second user's name and code shall be seized by the customer, and Right of customers shall be verify**;**

**Step 5:** the key is encrypted. Just after this step, if Valid key: Accept Starting Service cloud else Reject
*Customer g Of service provider*
**Step 6:** decrypt key by the service provider, the customer can access in his private space.
*Access to the space*
*Disconnect;*
*Disconnect;*

In this protocol, the user will have only to present his user name and the key in Step1 and step 5 to obtain a service.
Unlike other solutions this identification is encrypted, The encrypted and the decrypted will not be seen by the customer.

By this solution only user can be access to services and, it's better to encrypted all the data which will be transfer in the cloud, It sets a lot of time

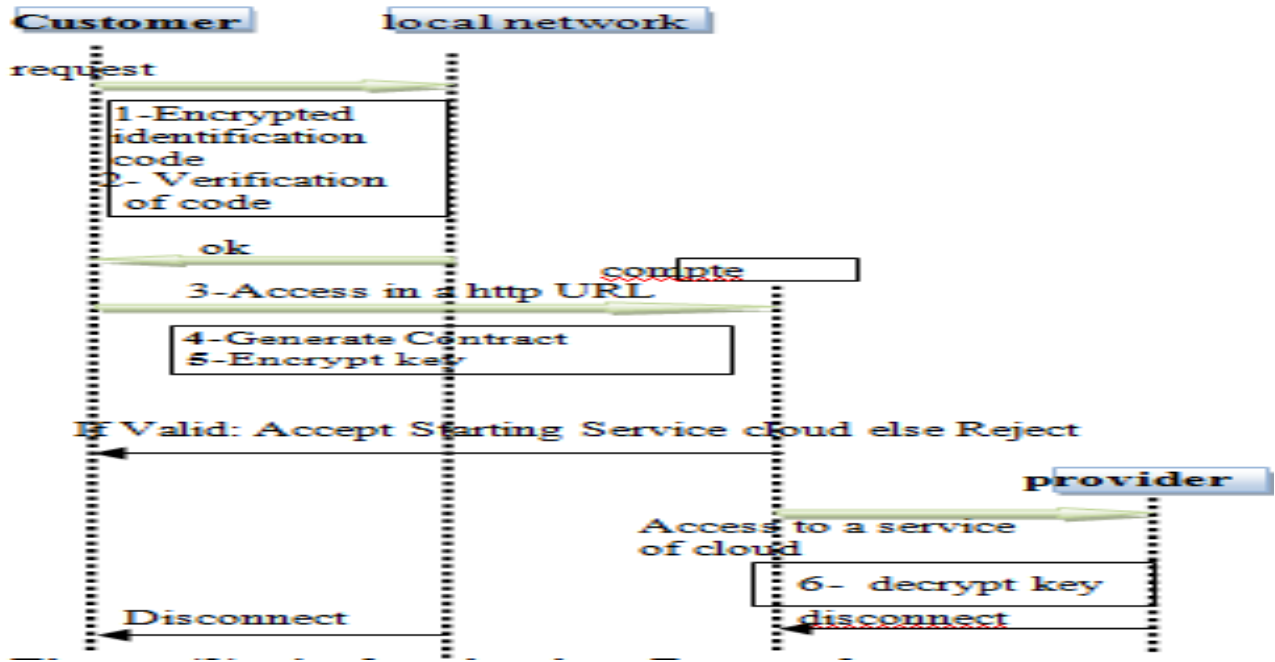

**Figure (3): Authentication Protocol**

CONCLUSION AND FUTURE WORKS

The cloud computing allows companies not only to protect their data and transform their spending investment into operational spending, But also to manage better their budget, because they pay generally only what they use. Numerous companies turn at present to Cloud computing for the saving, the archiving and the off-site restoration.

However, Cloud computing can be also used to have solutions of real time replication and high availability, To reduce at the most the interruptions of service and the losses of data. The companies which envisage the appeal to the cloud computing for the resumption after breakdown have to verify how their data, application and suppliers will handle diverse questions. Work is currently going on the frame work implantation where it will be applied to a specific case study. Further research could be realized to improve and to extend the present work.

REFERENCES

[1] R.Nithiavathy, (2015), *Data Integrity and Data Dynamics with Secure Storage Service in Cloud*, Proceedings of the 2015 International Conference on Pattern Recognition, Informatics and Mobile Engineering, IEEE,pp. 125-131, 2015.

[2] T.clement gagnon,The security in the cloud computing, , presented in CQSI ,October, 2012.

[3] R.Nithiavathy, (2013), *Data Integrity and Data Dynamics with Secure Storage Service in Cloud*, Proceedings of the 2013 International Conference on Pattern Recognition, Informatics and Mobile Engineering, IEEE,pp. 133-130, 2013.

[4] Mohammed Abdullatif Alzain and Eric Pardede, (2011), protection of data and cloud computing Europe septembre 2011.

[5] Alok Kumbhare, Yogesh Simmhan, Viktor Prasanna, (2016), *Cryptonite: A Secure and Performant Data* 2016 IEEE 5th International Conference on. IEEE,pp. 510-519, 2016.

[6] H. Huang and K. Liu, "Efficient key management for preserving HIPAA regulations," The Journal of Systems and Software, vol. 84, pp. 221–122, 2015.

[7] A. Gupta, A. Verma, P. Kalra, L. Kumar, "Big Data: A security compliance model", IT in Business, Industry and Government (CSIBIG), pp. 1 - 5, Indore, 2014.

[8] Cigref,cloud computing and protection of data, network companies entreprises 2015.

[9] H. Kamal Idrissi, A. Kartit, and M. El Marraki, "A taxonomy and survey of Cloud computing," presented at the Security Days (JNS3), 2013 National, 2013, pp. 1–5.

[10] R.C. Merkle,(1989),*A Certified Digital Signature*, Advances in Cryptology - CRYPTO '89, 9th Annual International Cryptology Conference, Santa Barbara, California, USA, Proceedings, Volume 435, pp. 218– 238, 1989.

[11] D. Benslimane, M. Barhamgi, F. Cuppens "PAIRSE: A Privacy- Preserving Service- Oriented Data Integration System", SIGMOD Record, (Vol. 42, No. 3), September 2013.

[12] Heffetz, Ori, and Katrina Ligett. Privacy and data-based research. No. w19433. National Bureau of Economic Research, 2013

[13] D.Deyan Chen, H. Zhao. "Data Security and Privacy Protection Issues in CC".IEEE International Conference on Computer Science and Electronics Engineering (ICCSEE).may 2015.

[14] A. Petit , Ben Mokhtar, L. Brunie and H. Kosch, "Towards Efficient and Accurate Privacy Preserving Web Search," MW4NG '14 December 8-12, 2014, Bordeaux, France. 2014.